%% file: template.tex
\documentclass{article}
\usepackage{dcase2019,amsmath,graphicx,url,times,booktabs, tabularx}
\usepackage[export]{adjustbox}
\usepackage{amssymb}
\usepackage{array}
\usepackage{booktabs}
\usepackage{algorithm, algorithmic}%
\usepackage{color}
\usepackage{subcaption}

\DeclareMathOperator{\LL}{\langle}
\DeclareMathOperator{\RR}{\rangle}

\title{A HYBRID PARAMETRIC-DEEP LEARNING APPROACH FOR SOUND EVENT LOCALIZATION AND DETECTION}

\name{Andr\'es P\'erez-L\'opez$^{1,2}$,
      Eduardo Fonseca$^{1}$\sthanks{E.F. is partially supported by a Google Faculty Research Award 2018.},
      Xavier Serra$^{1}$
      }
\address{$^1$ Music Technology Group, Universitat Pompeu Fabra, Barcelona\\  \{andres.perez, eduardo.fonseca, xavier.serra\}@upf.edu\\          
        $^2$ Eurecat, Centre Tecnol\`ogic de Catalunya, Barcelona\\
 }

\begin{document}

\ninept
\maketitle

\begin{sloppy}

\begin{abstract}
This work describes and discusses an algorithm submitted to the \textit{Sound Event Localization and Detection} Task of DCASE2019 Challenge.
The proposed methodology relies on parametric spatial audio analysis for source localization and detection, combined with a deep learning-based monophonic event classifier.
The evaluation of the proposed algorithm yields overall results comparable to the baseline system.
The main highlight is a reduction of the localization error on the evaluation dataset by a factor of 2.6, compared with the baseline performance.
\end{abstract}


\section{Introduction}
\label{sec:intro}

Sound Event Localization and Detection (SELD) refers to the problem of identifying, for each individual  event present in a sound field, the temporal activity, spatial location, and sound class to which it belongs. SELD is a current research topic which deals with microphone array processing and sound classification, with potential applications in the fields of signal enhancement, autonomous navigation, acoustic scene description or surveillance, among others.

SELD arises from the combination of two different problems: Sound Event Detection (SED) and Direction of Arrival (DOA) estimation. The number of works in the literature which jointly address SED and DOA problems is relatively small. It is possible to classify them by the type of microphone arrays used: distributed \cite{grobler2017sound, butko2011two, chakraborty2014sound} or near-coincident \cite{hirvonen2015classification, lopatka2016detection, Adavanne2018_JSTSP}.
As mentioned in \cite{Adavanne2018_JSTSP}, the usage of near-coincident circular/spherical arrays enables the representation of the sound field in the spatial domain, using the spherical harmonic decomposition, also known as Ambisonics \cite{gerzon1973periphony, daniel2000representation}. Such spatial representation allows a flexible, device-independent comparison between methods. Furthermore, the number of commercially available ambisonic microphones has increased in recent years due to their suitability for immersive multimedia applications.  
Taking advantage of the compact spatial representation provided by the spherical harmonic decomposition, several methods for parametric analysis of the sound field in the ambisonic domain have been proposed  \cite{pulkki2006directional, berge2010high, Politis2018, pulkki2018parametric}.
%These methods allow the sound field segmentation into direct and diffuse components, and further estimating the localization of the direct sounds.
These methods ease sound field segmentation into direct and diffuse components, and further localization of the direct sounds.
The advent of deep learning techniques for DOA estimation has also improved the results of traditional methods \cite{Adavanne2018_JSTSP}. However, none of the deep learning-based DOA estimation methods explicitly exploits the spatial parametric analysis. This situation is further extended to the SELD problem, with the exception of \cite{lopatka2016detection}, where DOAs are estimated from the \textit{active intensity vector}  \cite{pulkki2006directional}.

The motivation for the proposed methodology is two-fold.
First, we would like to check whether the usage of spatial parametric analysis in the ambisonic domain can improve the performance of SELD algorithms.
Second, temporal information derived by the parametric analysis could be further exploited to estimate event onsets and offsets, thus lightening the event classifier complexity; such reduction might positively impact algorithm's performance.

In what follows, we present the methodology and the architecture of the proposed system (Section~\ref{sec:method}). Then, we describe the design choices and the experimental setup (Section~\ref{sec:expe}), and discuss the results in the context of DCASE2019 Challenge - Task 3 (Section~\ref{sec:results}). A summary is presented in Section~\ref{sec:conclusion}. In order to support open access and reproducibility, all code is freely available at \cite{code}.

\section{Method}
\label{sec:method}

The proposed method presents a solution for the SELD problem splitting the task into four different problems: \textit{DOA  estimation}, \textit{association}, \textit{beamforming} and \textit{classification}, which will be described in the following subsections.
The former three systems follow a heuristic approach---in what follows, they will be jointly referred to as the \textit{parametric front-end}. Conversely, the \textit{classification} system is data-driven, and will be referred to as the \textit{deep learning back-end}. The method architecture is depicted in Figure~\ref{fig:diagram_all}.

\begin{figure}[h]
	\centering
    \includegraphics[width=\columnwidth]{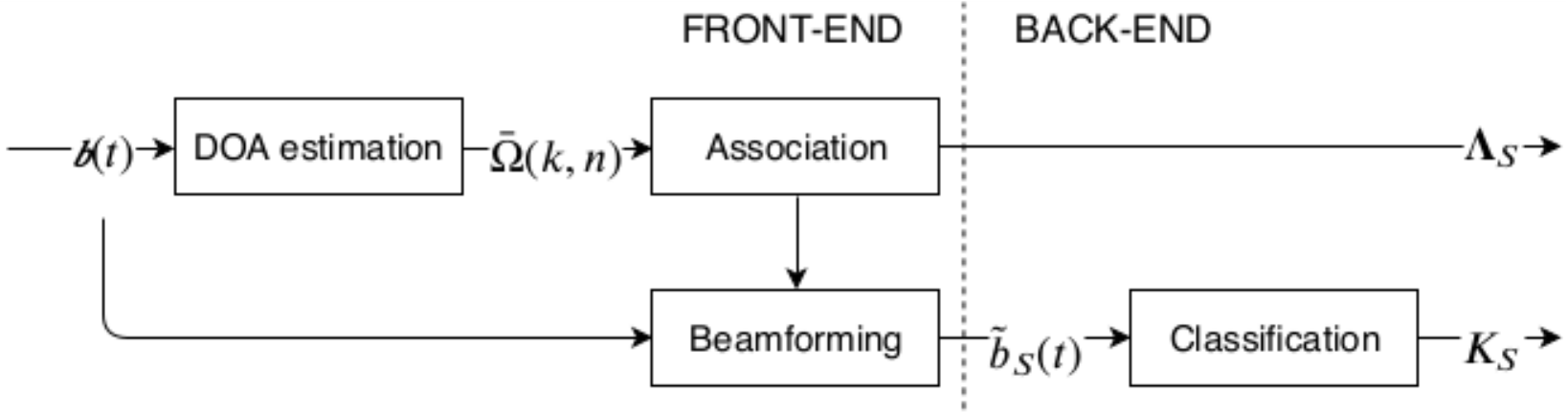}
    \caption{System architecture.}
    \label{fig:diagram_all}
    \vspace*{-4mm}
\end{figure}
% \vspace{-1mm}
\subsection{DOA estimation}
\label{ssec:doa_estimation}

\begin{figure*}[h]
    % \vspace{-2mm}
    \includegraphics[width=\textwidth]{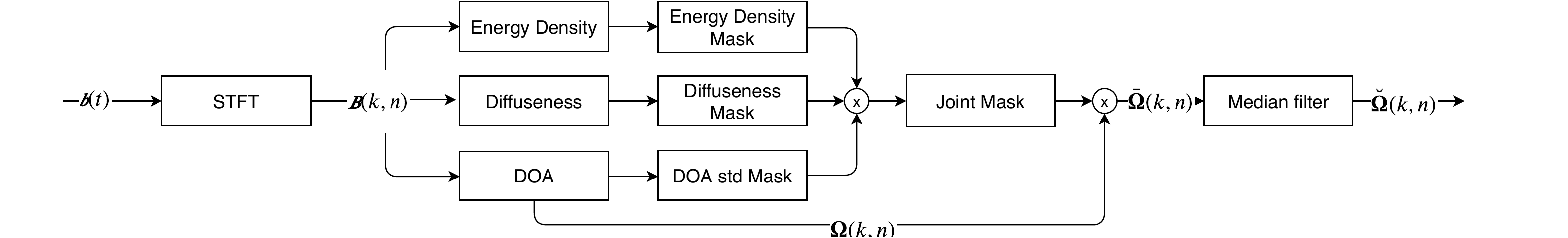}
    \caption{DOA estimation architecture.}
    \label{fig:doa}
    \vspace*{-5mm}
\end{figure*}

The \textit{DOA estimation} system (Figure~\ref{fig:doa}) is based on parametric time-frequency (TF) spatial audio analysis.
Let us consider a first-order ($L=1$) ambisonic signal vector $\pmb{b}(t)$ with N3D normalization \cite{carpentier2017normalization}:
\begin{equation}
    \pmb{b}(t) = [b_w(t), \sqrt{3}b_x(t),\sqrt{3}b_y(t), \sqrt{3}b_z(t)].
\end{equation}
From its short-time frequency domain representation $\pmb{B}(k,n)$, the instantaneous DOA at each TF bin $\pmb{\Omega}(k,n)$ can be estimated as:
\begin{equation}
    \begin{aligned}
    \pmb{I}(k,n) = -\frac{1}{Z_0}\mathbb{R}\{& [B_x(k,n), B_y(k,n), B_z(k,n)]  B_w(k,n)^* \},\\
    \pmb{\Omega}(k,n) &= [\varphi(k,n), \theta (k,n)] =\angle ( -\pmb{I}(k,n) ),
    \end{aligned}
\end{equation}
where $\pmb{I}(k,n)$ stands for the \textit{active intensity vector} \cite{pulkki2006directional}, $Z_0$ is the characteristic impedance of the medium, $^*$ represents the complex conjugate operator, and $\angle$ is the spherical coordinates angle operator, expressed in terms of azimuth $\varphi$ and elevation $\theta$.

It is desirable to identify the TF regions of $\pmb{\Omega}(k,n)$ which carry information from the sound events, and discard the rest. Three binary masks are computed with that aim.
%Next, with the aim of identifying the most relevant TF bins of $\pmb{\Omega}(k,n)$, three binary masks are computed.
% Next, we would like to filter $\pmb{\Omega}(k,n)$ in order to only include information from the bins where a sound event is present. For that goal we compute three different binary masks.
The first mask is the \textit{energy density mask}, which is used as an activity detector. The energy density $E(k,n)$ is defined as in \cite{stanzial1996reactive} :
\begin{equation}
    E(k,n) = \frac{|B_w(k,n)|^2
    + || [ B_x(k,n), B_y(k,n), B_z(k,n)] ||^2}{2Z_0c},
\end{equation}

with $c$ being the sound speed. A gaussian adaptive thresholding algorithm is then applied to $E(k,n)$, which selects TF bins with local maximum energy density, as expected from direct sounds.

The \textit{diffuseness mask} selects the TF bins with high energy propagation. Diffuseness $\Psi(k,n)$ is defined in \cite{merimaa2005spatial} as:
\begin{equation}
    \Psi(k,n) = 1 - ||\LL \pmb{I}(k,n) \RR|| / (c \LL E(k,n) \RR),
    \label{psi}
\end{equation}
where $\LL \cdot \RR$ represents the temporal expected value.

The third mask is the \textit{DOA variance mask}. It tries to select TF regions with small standard deviation\footnote{In this work, all statistical operators for angular position refer to the $2\pi$-\textit{periodic} operator for azimuth, and the standard operator for elevation.} with respect to their neighbor bins---a characteristic of sound fields with low diffuseness \cite{pulkki2018parametric}.

The three masks are then applied to the DOA estimation, obtaining the TF-filtered DOAs $\bar{\pmb{\Omega}}(k,n)$. Finally, a median filter is applied, with the aim of improving DOA estimation consistency and removing spurious TF bins.
The median filter is applied in a TF bin belonging to $\bar{\pmb{\Omega}}(k,n)$ only if the number of TF bins belonging to $\bar{\pmb{\Omega}}(k,n)$  in its vicinity is greater than a given threshold $B_{min}$.
The resulting filtered DOA estimation is referred to as $\breve{\pmb{\Omega}}(k,n)$.

\subsection{Association}
\label{ssec:association}

\begin{figure*}[h]
    \includegraphics[width=\textwidth]{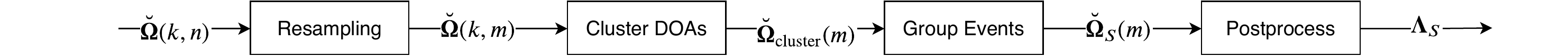}
    \caption{Association architecture.}
    \label{fig:association}
    \vspace*{-5mm}
\end{figure*}

The association step (Figure~\ref{fig:association}) tackles the problem of assigning the time-frequency-space observation $\breve{\pmb{\Omega}}(k,n)$ to a set of events, each one having a specific onset, offset  and location. 
% Note that the SED problem is still not completely solved at this point, since there is no event classification involved.
First, DOA estimates are resampled into \textit{frames} of the task's required length ($0.02$ s). In what follows, frames will be represented by index $m$. 
An additional constraint is applied: for a given window $n_0$, the DOA estimates $\breve{\pmb{\Omega}}(k,n_0)$ are assigned to the corresponding frame $m_0$ only if the number of estimates is greater than a threshold $K_{min}$.

Next, the standard deviation in azimuth ($\sigma_{\varphi}$) and elevation ($\sigma_{\theta} $) of the frame-based DOA estimates $\breve{\pmb{\Omega}}(k,m)$ are compared to a threshold value ($\sigma_{max}$), and the result is used to estimate the frame-based event overlapping amount $o(m)$ :
% based on the standard deviation in azimuth () and elevation () of the DOA estimates belonging to the 
%
%Next, a space-frequency clustering procedure is applied. For a given frame $m$, the standard deviation $\sigma$ of all DOA estimates $\breve{\pmb{\Omega}}(k,m)$ is computed, and the result is used to determine the overlapping amount $o(m)$
%using the following criterion:
\begin{equation}
\begin{aligned}
	o(m) =  \begin{cases}
    		1, &\text{if } \sigma_{\varphi}/2 + \sigma_{\theta} < \sigma_{max},\\
    		2, &\text{otherwise}.
    \end{cases}    		
\end{aligned}
\end{equation}
The clustered values $\pmb{\Omega}_{\text{cluster}}(m)$ are then computed as the $K=o(m)$ centroids of $\breve{\pmb{\Omega}}(k,m)$, using a modified version of K-Means which minimizes the central angle distance. Notice that, for $o(m)=1$, the operation is equivalent to the median.

The following step is the grouping of clustered DOA values into events.
Let us define $\pmb{\Omega}_S(m)$ as the frame-wise DOA estimations belonging to the event $S$. A given clustered DOA estimation $\pmb{\Omega}_{\text{cluster}}(m)$ belongs to the event $S$ if the following criteria are met:
\begin{itemize}
    \item The central angle between  $\pmb{\Omega}_{\text{cluster}}(m)$ and the median of $\pmb{\Omega}_S(m)$ is smaller than a given threshold $d_{max}^\text{ANGLE}$, and
    \item The frame distance between M and the closest frame of $\pmb{\Omega}_S(m)$ is smaller than a given threshold $d_{max}^\text{FRAME}$.
\end{itemize}

The resulting DOAs $\pmb{\Omega}_S(m)$ are subject to a postprocessing step with the purpose of 
delaying event onsets in frames where $o(m) > 2$, and discarding events shorter than a given minimum length. 
Finally, the frame-based event estimations are converted into \textit{metadata annotations} in the form $\pmb{\Lambda}_S = (\pmb{\Omega}_S, \text{onset}_S, \text{offset}_S)$.

%The resulting DOAs $\pmb{\Omega}_S(m)$ are subject to a postprocessing step, with the purpose of adjusting event activation times, and discarding events which are shorter than a given minimum length. 
%Finally, the frame-based event DOA estimations are converted into \textit{metadata annotations} in the form $\pmb{\Lambda}_S = (\pmb{\Omega}_S, \text{onset}_S, \text{offset}_S)$.

\subsection{Beamforming}
\label{ssec:Beamforming}

The last step performed in the front-end is the input signal segmentation. The spatial and temporal information provided by the annotations $\pmb{\Lambda}_S$ are used to produce monophonic signal estimations of the events, $\tilde{b}_S(t)$, as the signals captured by a virtual hypercardioid:
\begin{equation}
\tilde{b}_S(t) = \pmb{Y}(\pmb{\Omega}_S) \pmb{b}^\intercal(t),
\end{equation}
where $ \pmb{Y}(\pmb{\Omega}_S) = [Y_w(\pmb{\Omega}_S), Y_x(\pmb{\Omega}_S), Y_y(\pmb{\Omega}_S), Y_z(\pmb{\Omega}_S)]$ is the set of real-valued spherical harmonics up to order $L=1$ evaluated at $\pmb{\Omega}_S$.
%where $\pmb{Y}(\pmb{\Omega}_S)$ is the set of real-valued spherical harmonics up to order $L$ evaluated at the position $\pmb{\Omega}_S$, and $\pmb{\alpha}$ defines the virtual microphone directivity. In this work we have chosen a hypercardioid pattern, defined by $\pmb{\alpha} = [1, 1, 1, 1]^\intercal$.

\subsection{Deep learning classification back-end}
\label{ssec:backend_method}
\input{Backend_method}

%=================================================================================
%=================================================================================
%=================================================================================

\section{Experiments}
\label{sec:expe}

\subsection{Dataset, evaluation metrics and baseline system}
\label{ssec:dataset}
We use the TAU Spatial Sound Events 2019 - Ambisonic, which provides first-order ambisonic recordings.
%\footnote{Compatibility of the method with the \textit{Microphone dataset} could be straightforwardly accomplished with the usage of proper radial filters.}.
Details about the recording format and dataset specifications can be found in \cite{Adavanne2019_DCASE}.
The dataset features a vocabulary of 11 classes encompassing human sounds and sound events typically found in indoor office environments.
The dataset is split into a development and evaluation sets. 
The development set consists of a four fold cross-validation setup.

The SELD task is evaluated with individual metrics for SED (F-score (\textit{F}) and error rate (\textit{ER}) calculated in one-second segments) and DOA estimation (DOA error (\textit{DOA}) and frame recall (\textit{FR}) calculated frame-wise) \cite{Adavanne2018_JSTSP}.
%SED is evaluated with F-score (\textit{F}) and error rate (\textit{ER}) calculated in one-second segments, while DOA estimation is evaluated with two frame-wise metrics: DOA error (\textit{DOA}) and frame recall (\textit{FR}).
The \textit{SELD score} is an averaged summary of the system performance.

The baseline system features a CRNN that jointly performs DOA and SED through multi-task learning \cite{Adavanne2018_JSTSP}.
Baseline results are shown in Table \ref{tab:results_real}.

\subsection{Parametric front-end}
\label{ssec:param_frontend}

Based on the method's exploratory analysis, we propose the following set of parameter values, which are shown in Table~\ref{tab:params_front}.
% In general, the selected values follow a \textit{permissive} approach: most of parameters have relatively low values (e.g. $\Psi_{max}$, $K_{min}$, $\sigma_{max}$). The only exception is the median filter, which features a very large window size, and is responsible for TF filtering to a great extent. 

% The configuration could be considered as \textit{permissive} at the mask level, since the thresholding levels and minimum required amounts are in general small. This fact contrasts with the relatively big window of the median filter, which filters out the majority of bins. \textit{Association} parameters are also permissive

% \vspace*{-5mm}
\begin{table}[!htbp]
% \vspace{-3mm}
\centering
\caption{Parameter values for the selected configuration. Top: \textit{DOA analysis} parameters. Bottom: \textit{Association} parameters.}
\begin{tabular}{cccc}
\toprule
Parameter & Unit & Value\\
\midrule
% sampling rate & Hz  & 48000\\
STFT window size & sample  & 256\\
% STFT window overlap & sample & 128\\
% STFT window type & -  &  Hann\\
analysis frequency range & Hz & [0,8000]\\
time average vicinity radius $r$ & bin& 10 \\
diffuseness mask threshold $\Psi_{max}$ & - & 0.5 \\
energy density filter length & bin & 11\\
std mask vicinity radius & bin &  2\\
std mask normalized threshold & - & 0.15\\
median filter minimum ratio $B_{min}$ & - & 0.5 \\
median filter vicinity radius (k,n) & bin & (20, 20) \\
\midrule
% frame size $h$ & s  & 0.02\\
resampling minimum valid bins $K_{min}$ & bin & 1 \\
overlapping std threshold $\sigma_{max}$ & degree & 10 \\
grouping maximum angle $d_{max}^\text{ANGLE}$ & degree & 20\\
grouping maximum distance $d_{max}^\text{FRAME}$ & frame & 20\\
event minimum length & frame & 8\\
\bottomrule
\end{tabular}
\label{tab:params_front}
\vspace{-5mm}
\end{table}
% \vspace*{-5mm}

% \vspace{-1mm}
\subsection{Deep learning classification back-end}
\label{ssec:backend_experiments}
\input{Backend_experiments}

\section{Results and Discussion}
\label{sec:results}
\vspace{-2mm}
\begin{table}[!htbp]
\centering
% \caption{Evaluation results on development set (top) and evaluation set (bottom).}
\caption{Results for development (top) and evaluation (bottom) sets.}
\vspace{-2mm}
\begin{tabular}{cccccc}
\toprule
Method & \textit{ER} & \textit{F} & \textit{DOA} & \textit{FR} & \textbf{\textit{SELD}}\\
\midrule
Baseline & 0.34 & 79.9\% & 28.5$^{\circ}$  & 85.4\% & 0.2113\\
Proposed & 0.32 & 79.7\% & 9.1$^{\circ}$   & 76.4\% & \textbf{0.2026}\\
Ideal front-end & 0.08 & 93.2\% & $\sim0^{\circ}$   &$ \sim100$\% & 0.0379\\
\midrule
Baseline & 0.28 & 85.4\% & 24.6$^{\circ}$  & 85.7\% &  0.1764\\
Proposed & 0.29 & 82.1\% & 9.3$^{\circ}$   & 75.8\% & \textbf{ 0.1907}\\
\bottomrule
\end{tabular}
\label{tab:results_real}
\end{table}

Table~\ref{tab:results_real} shows the results of the proposed method for both development and evaluation sets, compared to the baseline.
Focusing on evaluation results, our method and the baseline obtain similar performance in SED (\textit{ER} and \textit{F}). However, there is a clear difference in the DOA metrics:
in our method, \textit{DOA} error is reduced by a factor of 2.6, but \textit{FR}  is $\sim10$ points worst.
In terms of \textit{SELD score}, our method performs slightly worse than the baseline in evaluation mode, while marginally outperforming it in development mode. 

The most relevant observation is the great improvement in \textit{DOA} error. Results suggest that using spatial audio parametric analysis as a preprocessing step can help to substantially improve localization. 
Figure \ref{fig:5a} provides further evidence for this argument: Challenge methods using some kind of parametric preprocessing (\textit{GCC-PHAT} with the microphone dataset, and \textit{Intensity Vector-Based} in ambisonics) obtained in average better \textit{DOA} error results.

\begin{figure}
  \begin{subfigure}[b]{\columnwidth}
    \includegraphics[width=\linewidth]{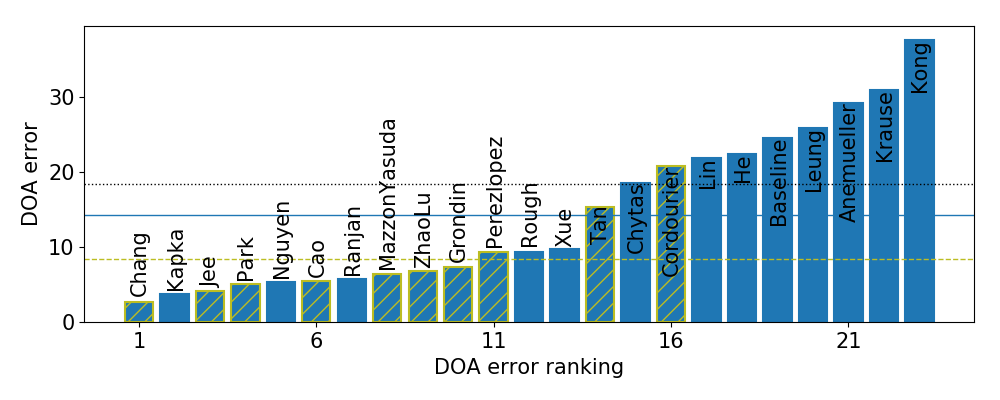}
    \vspace{-6mm}
    \caption{\textit{DOA error} across submissions. Hatched bars denote methods using parametric preprocessing. Horizontal lines depict average DOA error accross different subsets: all methods (solid), parametric methods (dashed), non-parametric methods (dotted).}
    \label{fig:5a}
  \end{subfigure}
  \begin{subfigure}[b]{\columnwidth}
    \includegraphics[width=\linewidth]{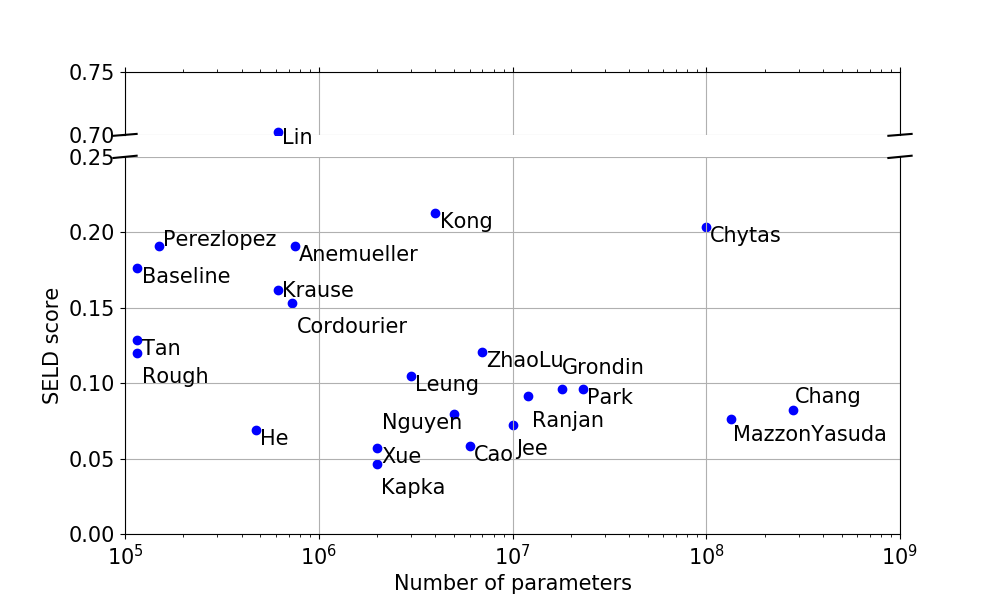}
    \vspace{-3mm}
    \caption{\textit{SELD score} versus complexity.}
    \label{fig:5b}
  \end{subfigure}
\caption{DCASE2019 Challenge Task 3 results, evaluation set.}
\vspace{-6mm}
\end{figure}

Conversely, the front-end fails regarding \textit{FR}. This is probably due to the complexity added by the association step \cite{Adavanne2018_JSTSP}, and its lack of robustness under highly reverberant scenarios. Including spectral information at the grouping stage might help to improve \textit{FR} --- such information could be provided by the classification back-end, in a similar approach to the baseline system. Another option would be the usage of more sophisticated source counting methods \cite{he2010detecting, Stefanakis2017}.

In order to gain a better insight of the classification back-end performance, Table~\ref{tab:results_real} shows the method results when the testing clips are obtained by feeding  the beamformer  with groundtruth annotations (\textit{ideal} front-end).
In this ideal scenario of DOA performance, the SED metrics show a significant boost. 
This result suggests that the low \textit{FR} given by the front-end has a severe impact on the back-end  performance.
Yet, the proposed system reaches similar performance to the baseline system in terms of SED metrics.

Finally, we would like to discuss algorithm complexity among Challenge methods. 
As depicted in Figure \ref{fig:5b}, there is a general trend towards architectures with very high number of weights, as a consequence of the usage of ensembles and large capacity networks.
Specifically, 66\% of submitted methods employ 1M weights or more, 30\% employ 10M or more, and 15\% employ 100M or more. Such complexities are several orders of magnitude greater than the baseline (150k weights) or the proposed method ($\sim$175k weights).
In this context, our method represents a low-complexity solution to the SELD problem, featuring a number of parameters and a performance comparable to the baseline method.

% \vspace*{-5mm}

%Average score for all fold data
%SELD score: 0.037888689433417076
%Total SED metrics: er: 0.08295702622686085, f:0.9316181038532054
%DOA metrics: doa error: 3.6480232543196293e-07, frame recall:0.9997841666666667
% \vspace*{-5mm}
\section{Conclusion}
\label{sec:conclusion}
% \vspace*{-2mm}
We present a novel approach for the SELD task. Our method relies on spatial parametric analysis for the computation of event DOAs, onsets and offsets. This information is used to filter the input signals in time and space, and the resulting event estimations are fed into a CRNN which predicts the class to which the events belong; the classification problem is thereby handled from a simple multi-class perspective.
The proposed method is able to obtain an overall performance comparable to the baseline system. 
The localization accuracy achieved by our method greatly improves the baseline performance, suggesting that spatial parametric analysis might enhance performance of SELD algorithms. Moreover, detection and classification performance in our method suffers from a low Frame Recall; improving this metric could lead to promising SELD scores.

%The localization performance is greatly improved, while the detection and classification performance suffers from the loss of Frame Recall. The improvement of this metric in the proposed system could lead to promising SELD scores, suggesting that spatial parametric analysis might lead to improve the classification task.

% -------------------------------------------------------------------------

\bibliographystyle{IEEEtran}
\bibliography{refs}

\end{sloppy}
\end{document}

%% file: Backend_method.tex
%intro - problem formulation==============
The parametric front-end performs DOA estimation, temporal activity detection and time/space segmentation, and produces monophonic estimations of the events, $\tilde{b}_S(t)$.
%The parametric front-end performs DOA estimation, temporal detection and time/space segmentation, the output being a monophonic signal containing, in theory, one single event.
Then, the back-end classifies the resulting signals as belonging to one of a target set of 11 classes.
Therefore, the multi-task nature of the front-end allows us to define the back-end classification task as a simple multi-class problem, even though the original SELD task is multi-label.
It must be noted, however, that due to the limited directivity of the first-order beamformer, the resulting monophonic signals can present a certain leakage from additional sound sources when two events overlap, even when the annotations $\pmb{\Lambda}_S$ are perfectly estimated.

%2 stages: context
%The classification method 
The classification method is divided into two stages.
First, the incoming signal is transformed into the log-mel spectrogram and split into TF patches.
%of $\mathbb{R}^{T\times F}$ (see Sec \ref{ssec:backend_experiments}). 
Then, the TF patches are fed into a single-mode based on a Convolutional Recurrent Neural Network (CRNN), which outputs probabilities for event classes $k \in \left\lbrace 1 ... K \right\rbrace$, with $K=11$.
%Then, we use a single model based on a Convolutional Recurrent Neural Network (CRNN), fed by the TF patches, and outputting probabilities for event classes $k \in \left\lbrace 1 ... K \right\rbrace$, with $K=11$.
Predictions are done at the event-level (not at the frame level), since the temporal activities have been already determined by the front-end.
%Only one label is predicted for each incoming event, such that there is no binarization stage needed as in a standard SED task.

\begin{figure}[ht]
  \vspace{-3mm}
  \centering
  \centerline{\includegraphics[width=0.84\columnwidth]{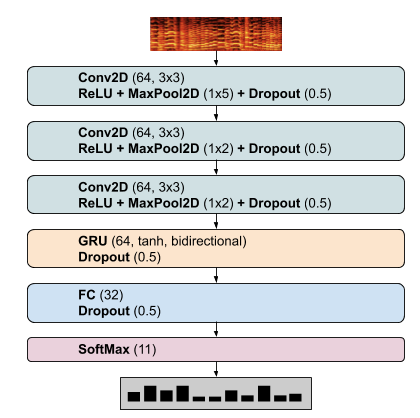}}
  \vspace{-2mm}
  \caption{Back-end architecture.}
  \label{fig:archi}
    \vspace{-5mm}
\end{figure}

%archi=============
The proposed CRNN is depicted in Figure \ref{fig:archi}.
It presents three convolutional \textit{blocks} to extract local features from the input representation.
Each convolutional block consists of one convolutional layer, after which the resulting feature maps are passed through a ReLU non-linearity \cite{nair2010rectified}.
This is followed by a max-pooling operation to downsample the feature maps and add invariance along the frequency dimension.
The target classes vary to a large extent in terms of their temporal dynamics, with some of them being rather impulsive (e.g., \textit{Door slam}), while others being more stationary (e.g., \textit{Phone ringing}). 
Therefore, after stacking the feature maps resulting from the convolutional blocks, this representation is fed into one bidirectional recurrent layer in order to model discriminative temporal structures.
%Specifically, 64 nodes of gated recurrent units (GRU) are used with \textit{tanh} activations.
The recurrent layer is followed by a Fully Connected (FC) layer, and finally a 11-way softmax classifier layer produces the event-level probabilities.
Dropout is applied extensively.
The loss function used is categorical cross-entropy.
The model has $\sim$175k weights.

%% file: Backend_experiments.tex
% training pipeline
 
We use the provided four fold cross-validation setup.
Training and validation stages use the outcome of an \textit{ideal} front-end, where the groundtruth DOA estimation and activation times are used to feed the beamformer for time-space segmentation.
Conversely, we test the trained models with the signals coming from the \textit{complete} front-end described in Section \ref{sec:method}.
%Model exploration: INTRO + Set overfitting as one of the main problems. Data scarcity.
We conducted a set of preliminary experiments with different types of networks including a VGG-like net, a less deep CNN \cite{Fonseca2019learning}, a Mobilenetv1 \cite{howard2017mobilenets} and a CRNN \cite{cakir2017convolutional}.
The latter was found to stand out, and we explore certain facets of the CRNN architecture and the learning pipeline.
%A number of decisions are taken to mitigate the risk of overfitting given data scarcity.

% pre-processing.
Sound events in the dataset last from $\sim$ 0.2 to 3.3 s.
First, clips shorter than 2s are replicated to meet this length.
Then, we compute TF patches of log-mel spectrograms of $T=50$ frames (1 s) and $F=64$ bands.
The values come from the exploration of $T \in \left\lbrace 25,50,75,100\right\rbrace$ and $F \in \left\lbrace 40,64,96,128\right\rbrace$.
%This is the result of exploration of $T \in \left\lbrace 25,50,75,100\right\rbrace$ and $F \in \left\lbrace 40,64,96,128\right\rbrace$.
$T=50$ is the top performing value, roughly coinciding with the median event duration. In turn, more than 64 bands provide inconsistent improvements, at the cost of increasing the number of network weights.

%architecture exploration
Several variants of the CRNN architecture were explored until reaching the network of Figure \ref{fig:archi}.
This included a small grid search over number of CNN filters, CNN filter size and shape, number of GRU units, number of FC units, dropout \cite{srivastava2014dropout}, learning rate, and the usage of Batch Normalization (BN) \cite{ioffe2015batch}.
Network extensions (involving more weights) were considered only if providing major improvements, as a measure against overfitting.
The main takeaways are: \textit{i)} squared 3x3 filters provide better results than larger filters, \textit{ii)} dropout of 0.5 is critical for overfitting mitigation, \textit{iii)} more than one recurrent layer does not yield improvements, while slowing down training, and \textit{iv)} surprisingly, slightly better performance is attained without BN nor pre-activation \cite{fonseca2018simple}.
% training------------------
For all experiments, the batch size was 100 and Adam optimizer was used \cite{kingma2014adam} with initial learning rate of 0.001, halved each time the validation accuracy plateaus for 5 epochs.
Earlystopping was adopted with a patience of 15 epochs, monitoring validation accuracy.
%inference
Prediction for every event was obtained by computing predictions at the patch level, and aggregating them with the geometric mean to produce a clip-level prediction.

% DA---------------------
Finally, we apply \textit{mixup} \cite{zhang2017mixup} as data augmentation technique.
Mixup consists in creating virtual training examples through linear interpolations in the feature space, assuming that they correspond to linear interpolations in the label space.
Essentially, virtual TF patches are created on the fly as convex combinations of the input training patches, with a hyper-parameter $\alpha$ controlling the interpolation strength.
Mixup has been proven successful for sound event classification, even in adverse conditions of corrupted labels \cite{Fonseca2019model}.
It seems appropriate for this task since the front-end outcome can present leakage due to overlapping sources, effectively mixing two sources while only one training label is available, which can be understood as a form of label noise \cite{Fonseca2019learning}.
Experiments revealed that mixup with $\alpha=0.1$ boosted testing accuracy in $\sim1.5$\%.

%% file: template.bbl
\begin{thebibliography}{10}
\providecommand{\url}[1]{#1}
\def\UrlFont{\rmfamily}
\providecommand{\newblock}{\relax}
\providecommand{\bibinfo}[2]{#2}
\providecommand\BIBentrySTDinterwordspacing{\spaceskip=0pt\relax}
\providecommand\BIBentryALTinterwordstretchfactor{4}
\providecommand\BIBentryALTinterwordspacing{\spaceskip=\fontdimen2\font plus
\BIBentryALTinterwordstretchfactor\fontdimen3\font minus
  \fontdimen4\font\relax}
\providecommand\BIBforeignlanguage[2]{{%
\expandafter\ifx\csname l@#1\endcsname\relax
\typeout{** WARNING: IEEEtran.bst: No hyphenation pattern has been}%
\typeout{** loaded for the language `#1'. Using the pattern for}%
\typeout{** the default language instead.}%
\else
\language=\csname l@#1\endcsname
\fi
#2}}

\bibitem{grobler2017sound}
C.~Grobler, C.~P. Kruger, B.~J. Silva, and G.~P. Hancke, ``Sound based
  localization and identification in industrial environments,'' in \emph{IECON
  2017-43rd Annual Conference of the IEEE Industrial Electronics
  Society}.\hskip 1em plus 0.5em minus 0.4em\relax IEEE, 2017, pp. 6119--6124.

\bibitem{butko2011two}
T.~Butko, F.~G. Pla, C.~Segura, C.~Nadeu, and J.~Hernando, ``Two-source
  acoustic event detection and localization: Online implementation in a
  smart-room,'' in \emph{2011 19th European Signal Processing
  Conference}.\hskip 1em plus 0.5em minus 0.4em\relax IEEE, 2011, pp.
  1317--1321.

\bibitem{chakraborty2014sound}
R.~Chakraborty and C.~Nadeu, ``Sound-model-based acoustic source localization
  using distributed microphone arrays,'' in \emph{2014 IEEE International
  Conference on Acoustics, Speech and Signal Processing (ICASSP)}.\hskip 1em
  plus 0.5em minus 0.4em\relax IEEE, 2014, pp. 619--623.

\bibitem{hirvonen2015classification}
T.~Hirvonen, ``Classification of spatial audio location and content using
  convolutional neural networks,'' in \emph{Audio Engineering Society
  Convention 138}.\hskip 1em plus 0.5em minus 0.4em\relax Audio Engineering
  Society, 2015.

\bibitem{lopatka2016detection}
K.~Lopatka, J.~Kotus, and A.~Czyzewski, ``Detection, classification and
  localization of acoustic events in the presence of background noise for
  acoustic surveillance of hazardous situations,'' \emph{Multimedia Tools and
  Applications}, vol.~75, no.~17, pp. 10\,407--10\,439, 2016.

\bibitem{Adavanne2018_JSTSP}
S.~Adavanne, A.~Politis, J.~Nikunen, and T.~Virtanen, ``Sound event
  localization and detection of overlapping sources using convolutional
  recurrent neural networks,'' \emph{IEEE Journal of Selected Topics in Signal
  Processing}, pp. 1--1, 2018.

\bibitem{gerzon1973periphony}
M.~A. Gerzon, ``Periphony: With-height sound reproduction,'' \emph{Journal of
  the Audio Engineering Society}, vol.~21, no.~1, pp. 2--10, 1973.

\bibitem{daniel2000representation}
J.~Daniel, ``Repr{\'e}sentation de champs acoustiques, application {\`a} la
  transmission et {\`a} la reproduction de sc{\`e}nes sonores complexes dans un
  contexte multim{\'e}dia,'' 2000.

\bibitem{pulkki2006directional}
V.~Pulkki, ``Directional audio coding in spatial sound reproduction and stereo
  upmixing,'' in \emph{Audio Engineering Society Conference: 28th International
  Conference: The Future of Audio Technology--Surround and Beyond}.\hskip 1em
  plus 0.5em minus 0.4em\relax Audio Engineering Society, 2006.

\bibitem{berge2010high}
S.~Berge and N.~Barrett, ``High angular resolution planewave expansion,'' in
  \emph{Proc. of the 2nd International Symposium on Ambisonics and Spherical
  Acoustics May}, 2010, pp. 6--7.

\bibitem{Politis2018}
A.~Politis, S.~Tervo, and V.~Pulkki, ``{COMPASS: Coding and Multidirectional
  Parameterization of Ambisonic Sound Scenes},'' \emph{IEEE International
  Conference on Acoustics, Speech, and Signal Processing}, no. May, pp.
  6802--6806, 2018.

\bibitem{pulkki2018parametric}
V.~Pulkki, S.~Delikaris-Manias, and A.~Politis, \emph{Parametric time-frequency
  domain spatial audio}.\hskip 1em plus 0.5em minus 0.4em\relax Wiley Online
  Library, 2018.

\bibitem{code}
\url{https://github.com/andresperezlopez/DCASE2019_task3}.

\bibitem{carpentier2017normalization}
T.~Carpentier, ``Normalization schemes in ambisonic: does it matter?'' in
  \emph{Audio Engineering Society Convention 142}.\hskip 1em plus 0.5em minus
  0.4em\relax Audio Engineering Society, 2017.

\bibitem{stanzial1996reactive}
D.~Stanzial, N.~Prodi, and G.~Schiffrer, ``Reactive acoustic intensity for
  general fields and energy polarization,'' \emph{The Journal of the Acoustical
  Society of America}, vol.~99, no.~4, pp. 1868--1876, 1996.

\bibitem{merimaa2005spatial}
J.~Merimaa and V.~Pulkki, ``Spatial impulse response rendering i: Analysis and
  synthesis,'' \emph{Journal of the Audio Engineering Society}, vol.~53,
  no.~12, pp. 1115--1127, 2005.

\bibitem{nair2010rectified}
V.~Nair and G.~E. Hinton, ``Rectified linear units improve restricted boltzmann
  machines,'' in \emph{Proceedings of the 27th international conference on
  machine learning (ICML-10)}, 2010, pp. 807--814.

\bibitem{Adavanne2019_DCASE}
\BIBentryALTinterwordspacing
S.~Adavanne, A.~Politis, and T.~Virtanen, ``A multi-room reverberant dataset
  for sound event localization and uetection,'' in \emph{Submitted to Detection
  and Classification of Acoustic Scenes and Events 2019 Workshop (DCASE2019)},
  2019. [Online]. Available: \url{https://arxiv.org/abs/1905.08546}
\BIBentrySTDinterwordspacing

\bibitem{Fonseca2019learning}
E.~Fonseca, M.~Plakal, D.~P.~W. Ellis, F.~Font, X.~Favory, and X.~Serra,
  ``Learning sound event classifiers from web audio with noisy labels,'' in
  \emph{Proc. IEEE ICASSP 2019}, Brighton, UK, 2019.

\bibitem{howard2017mobilenets}
A.~G. Howard, M.~Zhu, B.~Chen, D.~Kalenichenko, W.~Wang, T.~Weyand,
  M.~Andreetto, and H.~Adam, ``{MobileNets: Efficient Convolutional Neural
  Networks for Mobile Vision Applications},'' \emph{arXiv preprint
  arXiv:1704.04861}, 2017.

\bibitem{cakir2017convolutional}
E.~Cak{\i}r, G.~Parascandolo, T.~Heittola, H.~Huttunen, and T.~Virtanen,
  ``Convolutional recurrent neural networks for polyphonic sound event
  detection,'' \emph{IEEE/ACM Transactions on Audio, Speech, and Language
  Processing}, vol.~25, no.~6, pp. 1291--1303, 2017.

\bibitem{srivastava2014dropout}
N.~Srivastava, G.~Hinton, A.~Krizhevsky, I.~Sutskever, and R.~Salakhutdinov,
  ``Dropout: a simple way to prevent neural networks from overfitting,''
  \emph{The Journal of Machine Learning Research}, vol.~15, no.~1, pp.
  1929--1958, 2014.

\bibitem{ioffe2015batch}
S.~Ioffe and C.~Szegedy, ``Batch normalization: Accelerating deep network
  training by reducing internal covariate shift,'' in \emph{International
  Conference on Machine Learning}, 2015, pp. 448--456.

\bibitem{fonseca2018simple}
E.~Fonseca, R.~Gong, and X.~Serra, ``A simple fusion of deep and shallow
  learning for acoustic scene classification,'' in \emph{Proceedings of the
  15th Sound \& Music Computing Conference (SMC 2018)}, Limassol, Cyprus, 2018.

\bibitem{kingma2014adam}
\BIBentryALTinterwordspacing
D.~P. Kingma and J.~Ba, ``Adam: A method for stochastic optimization,'' in
  \emph{ICLR 2015}. [Online]. Available: \url{https://arxiv.org/abs/1412.6980}
\BIBentrySTDinterwordspacing

\bibitem{zhang2017mixup}
H.~Zhang, M.~Cisse, Y.~N. Dauphin, and D.~Lopez-Paz, ``mixup: Beyond empirical
  risk minimization,'' \emph{arXiv preprint arXiv:1710.09412}, 2017.

\bibitem{Fonseca2019model}
E.~Fonseca, F.~Font, and X.~Serra, ``Model-agnostic approaches to handling
  noisy labels when training sound event classifiers,'' in \emph{Proceedings of
  IEEE Workshop on Applications of Signal Processing to Audio and Acoustics},
  New York, US, 2019.

\bibitem{he2010detecting}
Z.~He, A.~Cichocki, S.~Xie, and K.~Choi, ``Detecting the number of clusters in
  n-way probabilistic clustering,'' \emph{IEEE Transactions on Pattern Analysis
  and Machine Intelligence}, vol.~32, no.~11, pp. 2006--2021, 2010.

\bibitem{Stefanakis2017}
N.~Stefanakis, D.~Pavlidi, and A.~Mouchtaris, ``{Perpendicular Cross-Spectra
  Fusion for Sound Source Localization with a Planar Microphone Array},''
  \emph{IEEE/ACM Transactions on Audio Speech and Language Processing},
  vol.~25, no.~9, pp. 1517--1531, 2017.

\end{thebibliography}
